\documentclass[11pt]{article}

\usepackage[margin=1in]{geometry}
\usepackage{setspace}        
\usepackage{hyperref}        
\usepackage{graphicx}        
\usepackage{booktabs}        
\usepackage{array}           
\usepackage{enumitem}        
\usepackage{float}
\usepackage{xcolor} 

\renewcommand{\textcolor}[2]{#2}

\usepackage{hyperref}
\hypersetup{hidelinks}

\pdfoutput=1
\usepackage[T1]{fontenc}
\usepackage[utf8]{inputenc}   
\usepackage{microtype}

\title{Accessibility Gaps in U.S. Government Dashboards for Blind and Low-Vision Residents}

\author{
    Chadani Acharya \\
    Texas A\&M University \\
    \texttt{chadani.acharya@tamu.edu} 
}

\date{} 

\begin{document}

\maketitle

\begin{abstract}
Public-facing dashboards have become the primary interface through which U.S. government agencies report high-stakes information to residents, including 
{\textcolor{red}{ current respiratory illness severity (CDC), homelessness and housing capacity (HUD), local housing production compliance (California HCD), city service performance (New York City’s Mayor’s Management Report), permit backlog and inspection timelines (Houston Permitting Center), and budget and public health status (City of Chicago)
\cite{cdcRespDashboard2025,hudPITDashboard2025,cahcdAPR2025,nycMMR2025,chicagoDashboards2025,houstonPermitting2025} }}. These dashboards are advertised as tools ``for the public,'' ``for customers,'' and ``a public report card,'' and several are updated weekly or even daily. Yet blind and low-vision residents who rely on screen readers may not be able to access this information independently, especially when critical values are only exposed through unlabeled charts, mouse-only hover tooltips, or visual “at-a-glance” cards with no text equivalent. {\textcolor{red}{ We audited six production dashboard ecosystems across federal, state, and municipal levels and evaluated each using an accessibility rubric grounded in screen reader needs and Web Content Accessibility Guidelines (WCAG)}}: {\textcolor{blue}{(1) discoverability of key metrics to assistive technology; (2) keyboard navigation without hover; (3) semantic labelling of axes, series, and categories; (4) presence of plain-language narrative summaries that explain current status and trend; and (5) availability of machine-readable tables or CSV downloads that mirror what sighted users see.}} {\textcolor{red}{ We find a recurring structural pattern we call \textit{urgency inversion}: the more time-critical and operational the dashboard (for example, Houston’s daily permit backlog dashboard), the less accessible it tends to be to blind and low-vision users, while slower accountability dashboards (for example, California’s annual housing compliance dashboard, New York City’s monthly performance “report card,” and CDC’s weekly respiratory severity summaries) are more likely to pair plain-language narrative with downloadable structured data.}} This inversion creates a civic participation gap: sighted residents receive immediate situational awareness and leverage over services, while blind and low-vision residents are pushed toward delayed channels, forced to reverse-engineer raw data, or required to ask staff directly, taking extra steps to achieve the same outcome. 
\end{abstract}


\section{Introduction}\label{sec:intro}

Across the United States, public agencies increasingly publish interactive ``open data dashboards'' to communicate government performance, risk, and accountability directly to the public. The U.S. Centers for Disease Control and Prevention (CDC) maintains weekly respiratory illness dashboards that classify national and state respiratory virus activity levels and trends for emergency department visits, framed explicitly as helping the public ``follow trends and understand the extent of respiratory illness activity'' in their community \cite{cdcRespDashboard2025}. The U.S. Department of Housing and Urban Development (HUD) operates Point{-}in{-}Time (PIT) and Housing Inventory Count (HIC) dashboards that let anyone explore homelessness counts and available shelter capacity by jurisdiction, subpopulation, and sheltered/unsheltered status \cite{hudPITDashboard2025}. The California Department of Housing and Community Development (HCD) publishes an Annual Progress Report (APR) dashboard that invites residents to ``track and download jurisdictions' progress toward their housing goals,'' translating legally required housing production and permitting reports from every city and county into a statewide accountability surface \cite{cahcdAPR2025}. Large cities now do the same at municipal scale: New York City’s Mayor’s Management Report (MMR) and Dynamic MMR describe themselves as a ``public report card on City services affecting New Yorkers,'' reporting thousands of agency performance indicators and historical trends \cite{nycMMR2025}; Chicago’s Office of Budget and Management advertises ``interactive public dashboards designed to make City data more accessible, engaging, and transparent for the public,'' including American Rescue Plan Act (ARPA) spending dashboards that link dollars to claimed outcomes using ``cleaned and validated City data'' \cite{chicagoDashboards2025}; and Houston’s Permitting Center publishes a performance dashboard ``designed for our customers'' that promises an ``at-a-glance view of our performance,'' including ``the total plans in the system for review and the number of business days of the oldest plan,'' and states that these metrics are updated daily \cite{houstonPermitting2025}. These descriptions are not neutral. When an agency calls a dashboard a ``public report card,'' ``designed for customers,'' or ``interactive public dashboards designed to make City data more accessible,'' it is making a civic promise: that residents will be able to understand what is happening.

For blind and low-vision (BLV) residents, who rely on screen readers and keyboard navigation instead of visual inspection, those promises often fail in practice~\cite{WebAIM-SR10-2024,WCAG21-Keyboard,Sharif-et-al-ASSETS-2021,Fan-et-al-TACCESS-2023}.
Accessibility research shows that dashboards and charts on the web frequently violate basic nonvisual requirements~\cite{Srinivasan-et-al-Azimuth-ASSETS-2023,Sharif-et-al-ASSETS-2021,Fan-et-al-TACCESS-2023,Thompson-et-al-ChartReader-CHI-2023}. 
{\textcolor{blue}{ First, many charts are not \emph{discoverable}: a screen reader lands only on an unlabeled canvas or hears ``graphic'' / ``chart object'' with no semantic structure, meaning the data may as well not exist \cite{Sharif2021ScreenReader}. Second, key values are often locked behind \emph{mouse-only hover} tooltips, which keyboard-only users and most screen reader users cannot trigger \cite{Sharif2021ScreenReader,Siu2021COVID}. Third, \emph{semantic labelling} is weak or missing: axes and series are not identified with meaningful text (for example, ``COVID-19 hospitalization rate per 100{,}000 adults 65+''), colour encodings like ``red = severe'' are not redundantly conveyed in text, and categories are sometimes only visually distinguished by colour \cite{Siu2021COVID,Fan2022TACCESS}. Fourth, BLV users repeatedly report that they need short \emph{narrative summaries in plain language} that answer ``what is happening now'' for e.g., ``hospitalizations increased this week,'' ``Very High activity in these regions,'' ``the oldest permit in review has waited 18 business days'' rather than being forced to reverse-engineer trends from a bar chart \cite{Siu2021COVID,Fan2022TACCESS,Siu2022AudioNarratives}. Finally, they need \emph{machine-readable tables or CSV downloads} that mirror the same core metrics shown visually, so they can examine exact numbers with assistive technology, instead of being shown an ``at-a-glance'' gauge that is not exposed to the accessibility tree \cite{Sharif2021ScreenReader,Fan2022TACCESS}.}} These expectations directly align with WCAG, the Web Content Accessibility Guidelines, which require perceivable, operable, understandable, and robust content, including providing text alternatives for non-text content, ensuring keyboard operability, avoiding color-only meaning, and supporting assistive technologies \cite{WCAG21,WCAG22}. A dashboard can be publicly online and visually polished, and still be effectively unusable to a BLV resident if it fails even one of these requirements.

The COVID-19 pandemic made this inequity visible in a brutal way. Studies of blind and low-vision users during the pandemic showed that many public COVID-19 dashboards and state health dashboards were unusable without sight: trend lines were encoded only as coloured curves on a chart, tooltips with case counts appeared only on mouse hover, and axes and markers were unlabeled or ambiguous to screen readers \cite{Siu2021COVID,Fan2022TACCESS}. Participants described spending large amounts of extra time trying to extract simple answers such as ``is it getting worse here'' or ``what is the current risk level in my county'' and, in many cases, having to ask a sighted friend for help despite these dashboards being advertised as public information \cite{Fan2022TACCESS}. Critically, these same studies documented a structural split between \emph{open data} and \emph{accessible narrative}: governments often considered their job done once they posted raw CSVs or interactive charts, but BLV users emphasized that a CSV is not the same as being told ``cases are peaking this week'' in plain language \cite{Siu2021COVID,Fan2022TACCESS}. Sighted users got instant trend messaging in the dashboard UI; BLV users got ``here is a dataset, interpret it ourself.'' Human--computer interaction (HCI) researchers responded by proposing new interaction techniques such as audio data narratives that automatically summarize major trends and outliers \cite{Siu2022AudioNarratives}, natural-language and voice interfaces that let a blind user ask a chart direct questions \cite{Alam2023SeeChart}, multimodal data representations combining speech, sonification, and tactile/braille exploration \cite{Seo2024MAIDR}, and smartphone-first navigation models like ChartA11y that reorganize visualizations into structured touch targets and spoken summaries for blind users \cite{Zhang2024ChartA11y}. This work argues that accessibility is not just exporting a CSV; it is designing an interaction model that delivers situational awareness and lets BLV users explore detail on demand.

At the same time, accessibility has shifted from ethics to law. In April 2024, the U.S. Department of Justice (DOJ) issued a final rule under Title~II of the Americans with Disabilities Act (ADA) that requires state and local governments to make their web content and mobile applications accessible \cite{DOJ2024TitleII,NLC2024DOJRule}. The DOJ rule explicitly adopts WCAG~2.1 Level~AA as the baseline technical standard and sets concrete compliance deadlines for public entities, with large jurisdictions on a roughly two-year clock and smaller jurisdictions on a roughly three-year clock from publication \cite{DOJ2024TitleII,NLC2024DOJRule,WCAG21}. In plain terms: city dashboards about housing approvals and permit backlog, state dashboards about local housing compliance, and city and federal dashboards about health risk and service performance are no longer just ``nice tools.'' They are regulated public services. Under Title~II, blind and low-vision residents must have equal opportunity to access and act on that information, not just eventual access via a phone call \cite{DOJ2024TitleII,NLC2024DOJRule}.

{\textcolor{red}{ However, two gaps remain in both research and practice. First, most existing empirical work either (a) audits health dashboards and crisis dashboards in isolation, primarily in the COVID-19 context \cite{Siu2021COVID,Fan2022TACCESS}, or (b) proposes novel accessible visualization techniques and interaction models that have not yet landed in production government dashboards \cite{Siu2022AudioNarratives,Alam2023SeeChart,Seo2024MAIDR,Zhang2024ChartA11y}. We do not yet have a cross-jurisdiction, multi-domain audit of the actual production dashboards that U.S. }} residents are told to use to track respiratory illness this week, homelessness and shelter capacity this year, housing permit compliance this cycle, agency performance ``dating back several years,'' and today’s permitting backlog in Houston \cite{cdcRespDashboard2025,hudPITDashboard2025,cahcdAPR2025,nycMMR2025,chicagoDashboards2025,houstonPermitting2025}. {\textcolor{red}{ Second, prior work has not explicitly connected accessibility failures to \emph{civic leverage}. }} Operational dashboards like Houston’s permitting dashboard claim to serve ``customers'' and update daily, reporting how long plans have been sitting in review \cite{houstonPermitting2025}. Yet those same dashboards often rely on hover-only graphics, visual “at-a-glance” cards, or interactive maps with no obvious machine-readable export, which makes them effectively inaccessible to BLV users. By contrast, slower dashboards like California’s annual housing compliance tracker, New York City’s performance ``public report card,'' Chicago’s budget/ARPA dashboards, and CDC’s weekly illness summaries are more likely to pair plain-language narrative summaries (``This Week’s Illness Severity Update,'' “progress toward housing goals,” “public report card on City services”) with downloadable structured data \cite{cahcdAPR2025,nycMMR2025,chicagoDashboards2025,cdcRespDashboard2025}. This recurring pattern suggests what we call \emph{urgency inversion}: the more time-sensitive and operational the information, the less accessible it tends to be to BLV residents.

This paper addresses these gaps. {\textcolor{red}{ We conduct an accessibility audit of six high-stakes public dashboard ecosystems spanning three levels of U.S. government (federal, state, and city) and multiple domains that directly affect daily life: national respiratory illness surveillance (CDC); federal homelessness and housing services (HUD PIT/HIC); state housing production accountability and enforcement (California HCD APR); and large-city performance dashboards in different U.S. regions and service areas (New York City’s Mayor’s Management Report, Chicago’s public performance/ARPA dashboards, and the Houston Permitting Center’s performance dashboard) \cite{cdcRespDashboard2025,hudPITDashboard2025,cahcdAPR2025,nycMMR2025,chicagoDashboards2025,houstonPermitting2025}.}}  {\textcolor{blue}{Our audit rubric operationalizes both WCAG requirements \cite{WCAG21,WCAG22} and empirically documented BLV needs \cite{Sharif2021ScreenReader,Siu2021COVID,Fan2022TACCESS,Siu2022AudioNarratives} into five concrete questions for each dashboard: (1) \emph{discoverability} of core metrics to a screen reader; (2) \emph{keyboard access} without mouse hover; (3) \emph{semantic labelling} of axes, series, and categories, including avoiding colour-only encodings; (4) \emph{presence of narrative summaries in plain language} that explain current status and change; and (5) \emph{availability of machine-readable tabular or CSV data} that mirrors what sighted users see. We also log each dashboard’s stated audience promise (``for residents,'' ``for customers,'' ``public report card'') and stated update cadence (daily, weekly, annual), because those claims determine how urgent the information is supposed to be and who the agency says it is serving \cite{nycMMR2025,chicagoDashboards2025,houstonPermitting2025}. }}

Our contributions are:
\begin{enumerate}
    \item A cross-government, multi-domain audit of six widely promoted U.S. public dashboards that now function as official interfaces to health risk, homelessness, housing enforcement, permitting backlog, agency performance, and local spending \cite{cdcRespDashboard2025,hudPITDashboard2025,cahcdAPR2025,nycMMR2025,chicagoDashboards2025,houstonPermitting2025}.
    \item An accessibility rubric that translates WCAG~2.1/2.2 criteria \cite{WCAG21,WCAG22} and BLV-reported pain points \cite{Sharif2021ScreenReader,Siu2021COVID,Fan2022TACCESS,Siu2022AudioNarratives} into concrete audit questions about discoverability, keyboard navigation, semantic labelling, narrative summaries, and machine-readable data access.
    \item Empirical evidence of \emph{urgency inversion}: dashboards that govern immediate, operational, high-stakes decisions (for example, ``how long is my permit delayed today'') are the least likely to provide accessible narrative or machine-readable data for BLV users, while slower accountability dashboards (for example, annual housing compliance or monthly performance ``report cards'') are more likely to provide both plain-language explanations and structured downloads \cite{cahcdAPR2025,nycMMR2025,chicagoDashboards2025,houstonPermitting2025,cdcRespDashboard2025}. We argue that this inversion is not cosmetic. It is an equity failure with direct implications for ADA Title~II compliance timelines \cite{DOJ2024TitleII,NLC2024DOJRule}.
\end{enumerate}

We frame accessibility of civic dashboards as a question of who gets timely situational awareness and leverage over government action.

\section{Methodology}













\subsection{Sampling Frame}
We selected six public dashboard ecosystems from United States government agencies at the federal, state, and municipal levels. Our goal was to capture systems that (a) directly affect residents’ daily safety, housing, or access to services, and (b) explicitly describe themselves as accountability or transparency tools for “the public,” “residents,” or “customers.” The six ecosystems are:

\textbf{CDC respiratory illness dashboards.} These weekly dashboards report current respiratory virus activity and severity in the United States, including COVID\textendash19, influenza, and RSV~\cite{cdcRespDashboard2025}. They provide state-level and demographic breakdowns and describe themselves as helping the public understand present illness risk.

\textbf{HUD homelessness dashboards.} These dashboards summarize the Point-in-Time Count and Housing Inventory Count data for each state and Continuum of Care region ~\cite{hudPITDashboard2025}. They report sheltered and unsheltered homelessness and vulnerable subpopulations such as families, veterans, and youth. They state that communities should use them to understand who is experiencing homelessness locally.

\textbf{California Department of Housing and Community Development Annual Progress Report (APR) dashboard.} This dashboard tracks how many housing applications were received, entitled, permitted, and completed by each jurisdiction in California each year, and positions itself as a way for the public to ``track and download jurisdictions’ progress toward their housing goals.~\cite{cahcdAPR2025}

\textbf{New York City Mayor’s Management Report (MMR) and Dynamic MMR.} The City of New York describes this as ``a public report card on City services affecting New Yorkers,'' offering comparative performance data across agencies ``dating back several years,'' with more than one thousand indicators updated monthly. These indicators are also published as an Open Data dataset.~\cite{nycMMR2025}

\textbf{Houston Permitting Center and Houston Public Works dashboards.} Houston publishes daily operational dashboards that are ``designed for customers,'' framed as ``an at-a-glance view of our performance,'' and described as ``updated daily.'' These dashboards report current backlog and review delays (for example, ``the number of business days of the oldest plan'' waiting in the queue) and provide infrastructure status views using interactive, typically map-based dashboards. ~\cite{houstonPermitting2025}

\textbf{Chicago budget / ARPA impact / respiratory illness dashboards.} Chicago describes its dashboards as ``interactive public dashboards designed to make City data more accessible,'' built on ``cleaned and validated City data.'' These include (a) weekly respiratory illness summaries for residents of Chicago and (b) financial/impact transparency around budget and ARPA spending.~\cite{chicagoDashboards2025}

Together, these six ecosystems cover public health risk, homelessness and housing services, state housing compliance and enforcement, city agency performance, permit and inspection backlog, and budget/spending impact. They also span different geographic contexts: national (CDC, HUD), state (California), and large cities in different U.S. regions (New York City, Houston, Chicago).  {\textcolor{violet}{ This diversity allows us to compare not just a single dashboard, but how different levels of government construct “public accountability.”}}

\subsection{Audit Rubric}
We evaluated each dashboard ecosystem using an audit rubric focused on accessibility for blind and low vision (BLV) residents who use screen readers.

\textbf{Screen reader discoverability.} We noted whether key content (for example, a chart, a metric card, or a status panel) appeared to be exposed as navigable text or structured elements, versus being embedded only in an unlabeled canvas, dynamic map tile, or hover-only tooltip. If a critical value existed only inside an interactive visualization with no textual fallback, we treated that as a discoverability failure.

\textbf{Keyboard navigation versus mouse-only hover.} We recorded whether a user could plausibly access the core values without relying on pointer hover. Dashboards that disclose numbers only on hover are effectively unusable for many BLV users and keyboard-only users.

\textbf{Semantic labelling.} We looked for meaningful textual labels on axes, legends, series, categories, and status indicators. Labels such as ``Very High respiratory activity,'' ``issued building permits,'' or ``number of business days of the oldest plan'' count as semantic. Labels such as ``chart object'' or colour-without-text do not.

\textbf{Plain-language narrative summaries.} We checked whether the dashboard provided a human-readable, short summary of current status and (ideally) direction of change. Examples include CDC’s “This Week’s Illness Severity Update,” Chicago’s weekly respiratory status summaries, or Houston’s statement of the current age of the oldest plan in review. We treated these summaries as essential because they let BLV users gain situational awareness without parsing a graphic.

\textbf{Machine-readable data access.} We captured whether the underlying values are available as accessible tables, CSV/Excel downloads, or open data portal entries. For example, New York City publishes its Mayor’s Management Report performance indicators as an Open Data dataset; California HCD publishes per-jurisdiction Annual Progress Report data statewide; HUD publishes Point-in-Time and Housing Inventory Count tables. Where dashboards did not clearly provide such access (for example, Houston’s daily backlog dashboard), we marked that as a gap.

\textbf{Temporal framing in text.} We noted whether the dashboard described change over time using words, such as “peaking nationally,” “remaining low,” “dating back several years,” “updated daily,” or “this week in Chicago.” Dashboards that only encode temporal change visually, without describing trend in text, require BLV users to reverse-engineer history from raw data.

\textbf{Stated audience / accountability promise.} We extracted how the agency itself framed the dashboard’s purpose. Some dashboards call themselves a “public report card,” others say they are “designed for customers,” and others urge residents to “track and download jurisdictions’ progress.” These promises matter because they define who is supposedly being served. We treat stronger public-facing rhetoric as implying a stronger ethical obligation to provide accessible access.

\subsection{Procedure and Analysis}
For each dashboard ecosystem, we reviewed its public-facing landing page, descriptive text, stated purpose, and any linked documentation or open data resources. We noted the stated update frequency (for example, ``updated daily,'' ``weekly summary,'' ``annual reporting,'' ``indicators updated monthly'') and recorded any explicit language describing current status, trend, or risk in plain language. All audits were conducted using publicly available dashboards and descriptions as accessed during the period of September 2025.

We then attempted to locate machine-readable versions of the dashboard’s core data: CSV downloads, accessible HTML tables with headers, open data portal datasets, or statewide jurisdiction-level spreadsheets. When such datasets were clearly linked as part of the dashboard’s presentation (for example, HUD’s Point-in-Time datasets on HUD Exchange; California HCD’s Annual Progress Report data; New York City’s Agency Performance Indicators in Open Data), we treated that as evidence of machine-readable access. When they were absent or unclear (for example, Houston’s permitting backlog dashboard and infrastructure status maps), we treated that as an accessibility gap.

{\textcolor{purple}{ Finally, we compared all six ecosystems across rubric dimensions and plotted them conceptually against two axes: (1) urgency of the information (daily operational status versus annual compliance reporting) and (2) accessibility affordances for blind and low vision users (presence of narrative summaries in text, presence of machine-readable data, presence of textual temporal framing). }} This comparison produced a consistent negative slope: as urgency rose, accessible affordances dropped. We refer to this recurring pattern as \textit{urgency inversion}.

Ethically, our analysis relies only on government dashboards and agency-published descriptions intended for the general public. We did not inspect nonpublic systems, did not collect personal data, and did not involve human subjects. This work therefore did not require institutional review board approval.

\begin{table}[H]
    \centering
    \scriptsize
    \begin{tabular}{p{2.8cm} p{1.6cm} p{2.4cm} p{2.4cm} p{2.4cm} p{2.8cm}}
    \toprule
    \textbf{Dashboard Ecosystem} &
    \textbf{Update cadence} &
    \textbf{Plain-language status / summary?} &
    \textbf{Downloadable / machine-readable data?} &
    \textbf{Trend explained in text?} &
    \textbf{Stated audience / promise} \\
    \midrule
    CDC (respiratory illness) &
    Weekly &
    Yes: ``This Week's Illness Severity Update'', categories like ``Very High'' &
    Yes: surveillance tables / datasets &
    Yes: ``peaking'', ``remaining low'' &
    ``Help the public follow trends'' \\
    \addlinespace
    HUD (homelessness PIT/HIC) &
    Annual &
    Limited: numeric counts exposed, little narrative of increase/decrease &
    Yes: PIT/HIC tables on HUD Exchange &
    Mostly visual; trend not narrated in text &
    ``Understand who experiences homelessness'' \\
    \addlinespace
    California HCD (APR) &
    Annual &
    Yes: explains permits issued as compliance benchmark &
    Yes: statewide per-jurisdiction dataset &
    Implied year-to-year &
    ``Track and download jurisdictions' progress'' \\
    \addlinespace
    NYC (MMR / Dynamic MMR) &
    Monthly / historical &
    Yes: framed as a ``public report card'' on agency performance &
    Yes: Agency Performance Indicators in Open Data &
    Yes: multi-year comparison ``dating back several years'' &
    ``Public report card'' / ``user friendly interface'' \\
    \addlinespace
    Houston (Permitting / Public Works) &
    Daily &
    Partial: snapshot like ``oldest plan in review has waited X business days'' &
    Not clearly offered publicly as CSV/table &
    No explicit ``getting better/worse'' narrative in text &
    ``Updated daily'', ``designed for customers'', ``at-a-glance'' \\
    \addlinespace
    Chicago (Budget / ARPA / Respiratory) &
    Weekly (health) / periodic (budget) &
    Yes: weekly respiratory summaries and ARPA/budget impact descriptions &
    Yes: ``cleaned and validated City data'' exposed publicly &
    Often yes: ``this week in Chicago'' / spending over time &
    ``Interactive public dashboards ... to make City data more accessible'' \\
    \bottomrule
    \end{tabular}
    \caption{Accessibility and accountability affordances across six U.S. public dashboard ecosystems. 
    Each row represents a dashboard group we audited. Columns describe: how fast it updates; whether it offers a direct plain-language status summary; whether the underlying data is available in machine-readable form (e.g., CSV, open data table); whether time trends are described in text rather than only shown visually; and how the dashboard frames its audience or purpose (e.g., ``public report card'', ``designed for customers''). 
    This table illustrates the pattern we call ``urgency inversion'': faster, operational dashboards (e.g., Houston, updated daily) are less likely to provide accessible narrative or downloadable data than slower dashboards (e.g., California HCD, NYC, CDC).}
    \label{tab:comparison}
\end{table}


\begin{figure}[H]
    \centering
    \includegraphics[width=\linewidth]{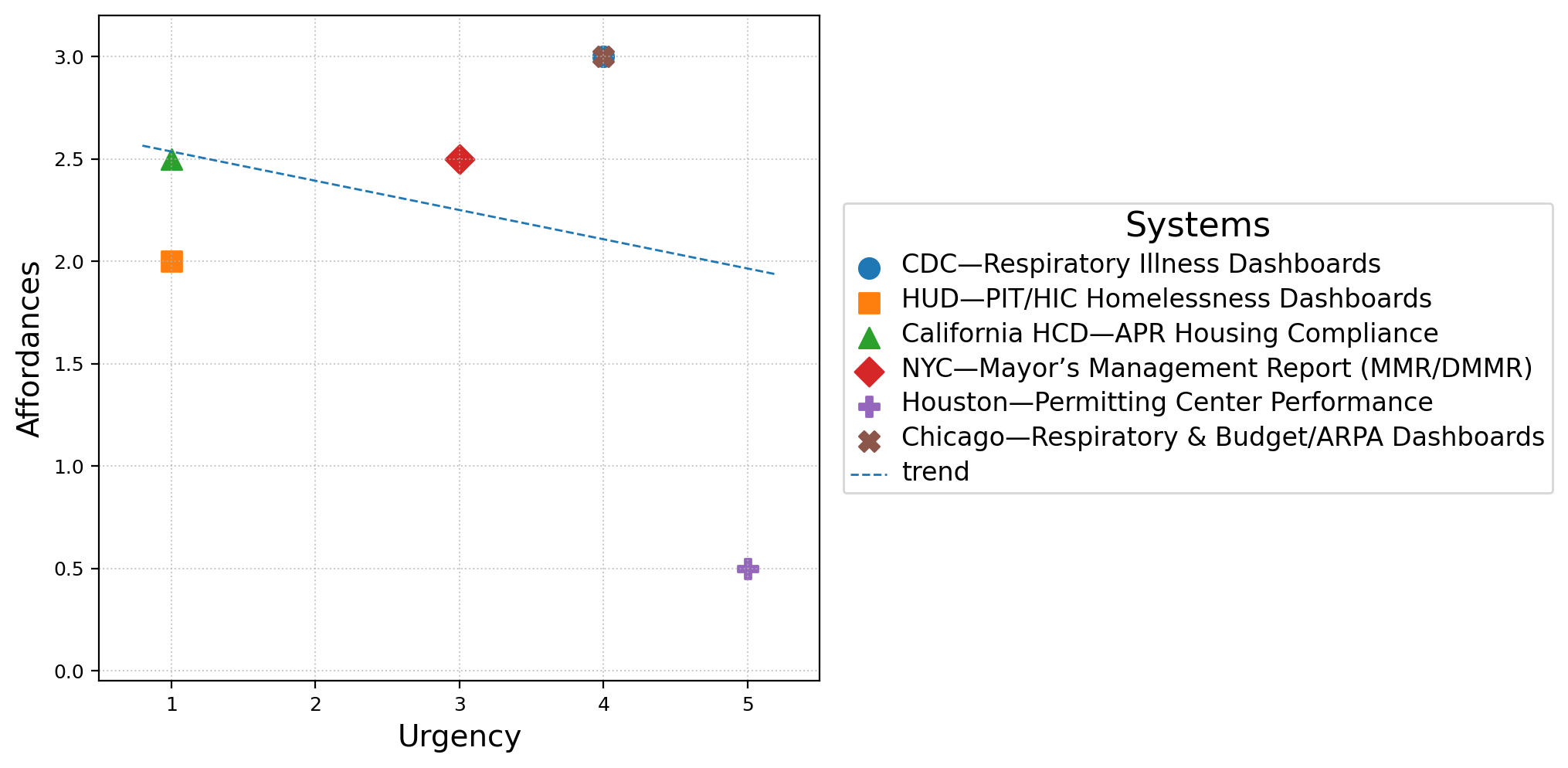}
    \caption{Accessibility affordances versus operational urgency across six public dashboard ecosystems. 
    Each marker represents one audited dashboard ecosystem: CDC respiratory illness dashboards; HUD PIT/HIC homelessness dashboards; California HCD Annual Progress Report (APR) housing compliance dashboard; New York City’s Mayor’s Management Report (MMR / Dynamic MMR); Houston Permitting Center performance dashboard; and Chicago’s respiratory illness and budget/ARPA dashboards. 
    The horizontal axis encodes how time-sensitive the information is, from annual compliance reporting (left) to daily backlog and inspection status (right). 
    The vertical axis encodes the strength of reported accessibility affordances for blind and low-vision (BLV) residents, computed as the sum of three features (each scored 0, 0.5, or 1): 
    (1) a plain-language status or “what is happening now” summary at the dashboard’s stated update cadence, 
    (2) clearly published machine-readable data (for example, CSV, open data table) that mirrors the same metrics shown visually, and 
    (3) explicit trend or change-over-time described in text rather than only in charts. 
    The dashed line shows the fitted linear trend. 
    The pattern illustrates \textit{urgency inversion}: dashboards with the most urgent, operational stakes (for example, Houston’s daily permitting backlog, “designed for customers” and “updated daily”) show the weakest accessible affordances for BLV users, while slower accountability dashboards (for example, California housing compliance, New York City’s performance “public report card,” CDC and Chicago’s weekly illness/risk summaries) provide both narrative context and downloadable data. 
    Audit snapshot: September~2025.}
    \label{fig:urgency_inversion}
\end{figure}


\begin{table}[H]
\centering
\small
\caption{\textbf{Contrasting communication styles across six U.S. public dashboard ecosystems.}
Each row presents a short ( $\leq$ 25 words) verbatim excerpt from the agency’s public-facing page or dashboard summary block describing what the dashboard offers to residents.
The excerpts illustrate two styles: (a) \emph{narrative status + trend} aimed at immediate public understanding (e.g., CDC, Chicago), and (b) \emph{operational/at-a-glance framing} that emphasizes update cadence or performance without explicitly narrating change or guaranteeing a data download (e.g., Houston; some accountability dashboards).
These quotes ground two rubric dimensions used in our audit and in Fig.~\ref{fig:urgency_inversion}: \emph{narrative\_now} (a plain-language “what’s happening now” statement) and \emph{trend\_text} (an explicit textual description of change over time).
Ellipses (\ldots) indicate brief omissions to keep within fair use; punctuation is normalized for line breaking; citations point to the official sources as accessed on 9 September 2025.}
\label{tab:communication_styles}

\vspace{4pt}
\begin{tabular}{p{3cm} p{11cm}}
\toprule
\textbf{System} & \textbf{Excerpt} \\
\midrule
CDC (weekly) & “Hospitalization rates for COVID-19 are peaking nationally, while rates for influenza and RSV remain low.” \cite{cdcRespDashboard2025} \\
HUD (annual) & “The reports contain summary data from the Point-in-Time (PIT) count and Housing Inventory Count (HIC).” \cite{hudPITDashboard2025} \\
California HCD (annual) & “The key benchmark for tracking RHNA progress is issued building permits.” \cite{cahcdAPR2025} \\
NYC MMR (monthly) & “\ldots serves as a public report card on City services affecting New Yorkers.” \cite{nycMMR2025} \\
Houston (daily) & “Updated daily, this dashboard was designed for customers and provides an at-a-glance view of our performance.” \cite{houstonPermitting2025} \\
Chicago (weekly) & “The Respiratory Illness Dashboard \ldots summarizes information about respiratory virus disease activity in Chicago.” \cite{chicagoDashboards2025} \\
\bottomrule
\end{tabular}

\vspace{4pt}
\footnotesize\emph{Note.} Quotes are chosen to represent how each ecosystem communicates current status to residents; they are not exhaustive of all on-page text. These lines map to our audit features \emph{narrative\_now} and \emph{trend\_text} and motivate the differences visualized in Fig.~\ref{fig:urgency_inversion}.
\end{table}

\section{Results}

\subsection{CDC Respiratory Illness Dashboards}
The Centers for Disease Control and Prevention (CDC) publishes weekly respiratory illness dashboards that report national and state-level severity for respiratory viruses such as COVID\textendash19, influenza, and RSV. The stated purpose is to help the public ``follow trends and understand the extent of respiratory illness activity'' nationally and locally, rather than to target only clinicians or researchers. The dashboards present a recurring plain-language block such as ``This Week's Illness Severity Update,'' which labels current activity levels using human-readable categories (for example, ``Very High'' or ``Low'') and explicitly describes whether hospitalization rates are increasing, stable, or decreasing. The CDC also exposes the underlying surveillance data in structured, machine-readable form: hospitalization rates per age group, per virus, per state, and similar epidemiological indicators are typically available as tabular data. Importantly, time is narrated in text, not only drawn as a line. The dashboard uses wording such as ``peaking nationally'' or ``remaining low this week,'' which means that a blind or low vision resident using a screen reader can get immediate situational awareness without having to interpret a color-coded chart. In terms of our audit rubric, CDC provides (1) a plain-language narrative summary of current risk, (2) explicit temporal framing in words, and (3) machine-readable data tables that mirror the visualized values.

\subsection{HUD Homelessness Dashboards}
The U.S. Department of Housing and Urban Development (HUD) publishes homelessness dashboards derived from the annual Point-in-Time (PIT) Count and Housing Inventory Count (HIC). These dashboards report, for each state and Continuum of Care region, the number of people experiencing homelessness, disaggregated by sheltered versus unsheltered status and by subpopulations such as families with children, veterans, and youth. The stated purpose is to help communities understand who is experiencing homelessness in their area and what level of assistance is needed. The core data behind these dashboards is publicly downloadable via HUD Exchange as machine-readable tables. Charts on the dashboard are explicitly described as being generated from those publicly posted datasets.

From an accessibility perspective, this makes the raw numbers available to a blind or low vision user. However, the dashboards do not consistently provide a short, human-readable trend summary in text (for instance, ``unsheltered youth homelessness increased in this Continuum of Care compared to last year''). Instead, those trends are largely expressed through bar charts across years or side-by-side comparisons, which typically require visual inspection or mouse hover. In our rubric terms, HUD strongly supports machine-readable data access, but the narrative summary and temporal framing are weak: a blind user can access the counts, but must perform their own analysis to determine change over time, urgency, or inequity. Sighted users, by contrast, can infer those shifts at a glance from the visuals.

\subsection{California Housing and Community Development Annual Progress Report Dashboard}
The California Department of Housing and Community Development (HCD) maintains a statewide Annual Progress Report (APR) dashboard that tracks local housing production and compliance with state housing law. Every jurisdiction in California is legally required to submit an APR each year, reporting how many housing applications were received, how many projects were entitled, how many permits were issued, and how many units were completed. HCD’s dashboard tells the public to ``track and download jurisdictions' progress toward their housing goals,'' and it clearly defines the main accountability metric: issued building permits toward each jurisdiction’s required housing targets.

This is important for accessibility because the dashboard does not only expose numbers, it also teaches interpretation. By stating that permits issued are the enforcement benchmark, the state effectively supplies a lens: this is how to tell whether our city is doing its job. HCD also distributes the APR data as a statewide, per-jurisdiction dataset that can be downloaded in machine-readable form. Performance is organized by jurisdiction and year, so progress over time is embedded in the data model, even if not always narrated sentence-by-sentence. In our rubric, California’s dashboard provides a clear audience promise (public oversight of compliance), machine-readable data for every jurisdiction, and a plain-language framing of what “progress” means. This lowers the interpretive burden for blind and low vision users, because they do not have to guess which variable matters.

\subsection{New York City Mayor’s Management Report and Dynamic MMR}
New York City’s Mayor's Management Report (MMR) and its interactive companion, the Dynamic MMR, are explicitly framed as ``a public report card on City services affecting New Yorkers.'' The City describes these dashboards as a user-friendly way to review comparative performance data across all agencies, and emphasizes that the indicators date back several years. The scope is broad: the City tracks thousands of indicators, with more than 1{,}000 reported monthly, capturing dimensions such as service quality, timeliness, cleanliness, safety, and more. Critically, New York City also publishes the Agency Performance Indicators dataset in its Open Data portal. That dataset includes indicator names, values, time stamps, and responsible agencies.

From an accessibility standpoint, this pairing of narrative framing (``public report card'') with a machine-readable dataset is powerful. A blind or low vision resident who uses a screen reader can, in principle, download the same performance indicators that sighted residents view through charts. Because the City also describes trends historically (``dating back several years''), the notion of change over time is not purely visual. In our rubric, New York City strongly meets the criteria of stated accountability promise, temporal framing in text, and machine-readable data. The City is essentially treating accessibility as part of civic oversight: we are supposed to be able to check whether our agencies are failing you.

\subsection{Houston Permitting Center and Public Works Dashboards}
Houston publishes operational dashboards through its Permitting Center and Public Works that are explicitly described as ``designed for customers,'' ``updated daily,'' and providing an ``at-a-glance view of our performance.'' These dashboards report, for example, ``the total plans in the system for review and the number of business days of the oldest plan,'' as well as infrastructure status through ArcGIS-style maps intended to help residents navigate ongoing work. This is the most time-sensitive information in our sample: it directly affects whether a permit application is stalled right now, how long an applicant may have to wait, and what the oldest unreviewed plan in the queue looks like today.

In accessibility terms, Houston is also the point of greatest concern. The dashboards emphasize immediate visual status and map-based interaction, but we did not find clear evidence of downloadable CSVs or accessible HTML tables that expose the same core backlog metrics. We also did not see consistent textual narration of trend (for example, ``backlog has improved since last week''). Instead, the dashboards communicate “today’s snapshot,” expecting that the user can perceive on-screen graphics. For a sighted contractor or homeowner, this provides instant leverage in dealing with the city. For a blind contractor or homeowner, the same leverage likely requires phoning staff or sending email, because the dashboard does not explicitly guarantee a screen-reader-friendly way to inspect the numbers. In our rubric, Houston has the highest urgency (daily updates, direct service impact), but the weakest visible support for narrative summaries, machine-readable data, and temporal framing in text.

\subsection{Chicago Budget, ARPA Impact, and Respiratory Illness Dashboards}
The City of Chicago hosts multiple public dashboards that it describes as ``interactive public dashboards designed to make City data more accessible,'' built on ``cleaned and validated City data.'' Two categories matter for this study. First, Chicago’s public health / respiratory illness dashboard provides weekly summaries of COVID-19, influenza, and RSV activity in the city. These summaries describe circulating strains and current risk levels in plain language, which gives residents immediate situational awareness without requiring them to interpret a chart. Second, Chicago’s budget and ARPA (American Rescue Plan Act) impact dashboards are designed to show where public recovery dollars are being spent and what impacts are being claimed in terms of outcomes or service reach.

Chicago also publicly ties these dashboards to downloadable or explorable structured data. The language about ``cleaned and validated City data'' signals that residents are expected to be able to inspect the numbers, not just view a static infographic. The respiratory illness dashboard mirrors CDC’s approach by offering a “this week in Chicago” narrative, while the ARPA/budget dashboards frame financial transparency and claimed impact over time. As in New York City, this pairing of narrative plus machine-readable data suggests intentional accessibility. However, certain infrastructure and neighborhood-level maps (like those often published through ArcGIS) likely still rely on hover-based, visually encoded spatial layers. This means neighborhood-scale equity questions may remain less accessible to blind and low vision users than citywide summary statements.

\subsection{Cross-System Pattern: Urgency Inversion}
Comparing all six ecosystems reveals a structural pattern we call \textit{urgency inversion}. Dashboards with the most time-sensitive, operational stakes such as Houston’s daily permit backlog and infrastructure status dashboards, which directly affect whether a resident or contractor can move forward today are the least explicitly accessible to blind and low vision users. They emphasize being ``updated daily,'' ``designed for customers,'' and ``at-a-glance,'' but they do not consistently provide machine-readable tables, trend descriptions in text, or an alternative to visually inspecting interactive maps.

By contrast, dashboards that update more slowly and operate at the level of compliance, accountability, or policy oversight (for example, California’s Annual Progress Report housing dashboard, New York City’s Mayor’s Management Report, Chicago’s budget/ARPA dashboards, and CDC’s weekly respiratory illness summaries) tend to provide one or both of two affordances: (1) plain-language narrative that explains what is happening right now and how it is changing, and (2) downloadable structured data (CSV, open data tables, per-jurisdiction datasets) that blind and low vision users can consume with assistive technology. 

In practice, this means sighted residents get instant operational leverage exactly where timing matters most, while blind and low vision residents are pushed toward slower, policy-scale dashboards or forced to self-advocate to access urgent information. Urgency, which should increase the ethical demand for accessibility, instead correlates with weaker accessible affordances.

\section{Discussion}
\label{sec:discussion}

\subsection{Accountability Without Accessibility}
Public agencies now describe their dashboards using explicitly civic language. New York City calls the Mayor’s Management Report (MMR) a ``public report card on City services affecting New Yorkers'' and emphasizes transparency and accountability across thousands of indicators \cite{nycMMR2025}. California’s Department of Housing and Community Development (HCD) invites residents to ``track and download jurisdictions' progress toward their housing goals,'' framing housing production as something the public should be able to monitor and enforce \cite{cahcdAPR2025}. Chicago’s Office of Budget and Management advertises ``interactive public dashboards designed to make City data more accessible, engaging, and transparent for the public,'' built on ``cleaned and validated City data'' \cite{chicagoDashboards2025}. The CDC tells residents that its respiratory dashboards are there to help ``the public follow trends and understand the extent of respiratory illness activity'' in their community, week by week \cite{cdcRespDashboard2025}. Houston’s Permitting Center promises an ``at-a-glance view of our performance,'' ``designed for our customers,'' and says metrics are ``updated daily'' so applicants can see ``the total plans in the system for review and the number of business days of the oldest plan'' \cite{houstonPermitting2025}.

All of that language is a promise to residents: we will tell you what is happening. What none of these descriptions do is name blind and low-vision (BLV) residents explicitly, or assert that the information is available to screen reader users on equal footing. The dashboards invoke ``the public,'' ``customers,'' and ``transparency'' without committing to WCAG conformance, keyboard-only operability, or nonvisual discoverability of charts \cite{WCAG21,WCAG22}. In practice, that means the phrase ``public report card'' often still assumes a sighted public. This is not a semantic quibble. Under Title~II of the Americans with Disabilities Act (ADA), as updated by the U.S. Department of Justice’s 2024 rule, state and local governments are now required to provide equal access to digital services, and WCAG~2.1~AA is the governing standard \cite{DOJ2024TitleII,NLC2024DOJRule,WCAG21}. When a city says ``updated daily for our customers,'' that is no longer marketing copy. It is a civil rights claim.

\subsection{Accessible Data vs.\ Accessible Interpretation}
Several dashboards in our sample expose raw numbers in machine-readable form. HUD’s homelessness dashboards point to the Point{-}in{-}Time (PIT) and Housing Inventory Count (HIC) tables, which are downloadable as structured data by geography and subpopulation \cite{hudPITDashboard2025}. California HCD publishes per-jurisdiction Annual Progress Report (APR) data and explicitly encourages the public to download and compare local housing performance \cite{cahcdAPR2025}. New York City publishes the Mayor’s Management Report indicators to the City’s Open Data portal, including agency, metric, timestamp, and value \cite{nycMMR2025}. Chicago emphasizes that its ARPA/budget dashboards are backed by ``cleaned and validated City data,'' implying that the underlying numbers are intended to be inspectable rather than decorative \cite{chicagoDashboards2025}.

Providing machine-readable data is necessary, but we find that it is not sufficient for equal access. Prior work on BLV access to COVID-19 dashboards and other civic data showed that telling a blind user ``the CSV is available'' is not the same as telling them ``hospitalizations increased this week and risk is high'' \cite{Siu2021COVID,Fan2022TACCESS}. Structured data offloads interpretation to the user. Sighted residents hovering over a chart get instant narrative framing: red means danger, the curve is rising, backlog is getting worse. BLV residents are often handed a spreadsheet and told, implicitly, to build their own analysis pipeline. That is extra cognitive labor, extra time, and in many cases extra social labor (asking a sighted coworker or family member for help) \cite{Siu2021COVID,Fan2022TACCESS,Sharif2021ScreenReader}.

Some dashboards in our audit already move beyond ``here is data'' toward ``here is the situation.'' The CDC’s respiratory illness dashboards publish a recurring plain-language block like ``This Week’s Illness Severity Update,'' using categorical terms such as ``Very High'' or ``Low,'' and explicitly describe whether hospitalization rates are increasing or stabilizing \cite{cdcRespDashboard2025}. Chicago’s respiratory illness dashboard similarly summarizes ``this week in Chicago,'' describing circulating strains and local burden in text, rather than assuming a user can visually interpret a small-multiples chart \cite{chicagoDashboards2025}. These narrative summaries directly address what BLV participants in prior studies requested: immediate situational awareness without sight, plus the option to drill into numbers later \cite{Siu2021COVID,Siu2022AudioNarratives}. By contrast, HUD’s PIT/HIC dashboards and many housing/permit backlogs present counts and trends primarily through visual comparisons and bar charts, with little or no textual narration of trend or urgency. A blind user can download the tables, but will not get ``homelessness among unsheltered youth increased in our Continuum of Care'' spelled out in words.

The result is a split between \emph{accessible data} and \emph{accessible interpretation}. A dashboard that only publishes CSVs satisfies open data culture but fails people who need to act right now. A dashboard that only publishes visual cards satisfies sighted intuition but fails assistive technology. The equitable pattern is both: a short, plain-language narrative of ``what is happening now,'' updated at the same cadence as the dashboard, and a machine-readable table with the exact same metrics \cite{Siu2021COVID,Fan2022TACCESS,Siu2022AudioNarratives,WCAG21,WCAG22}.

\subsection{Urgency Inversion as Structural Inequity}
Our central finding is what we call \emph{urgency inversion}. The more time-sensitive and operational a dashboard’s information is, the less accessible it tends to be to BLV residents.

Houston’s permitting dashboard is the clearest case. It is described as ``designed for our customers'' and ``updated daily,'' and it surfaces metrics like ``the total plans in the system for review and the number of business days of the oldest plan'' information that directly affects whether a builder or homeowner can proceed with work this week \cite{houstonPermitting2025}. Yet the dashboard’s presentation leans on visual, at-a-glance cards and interactive/ArcGIS-style elements. We did not find clear evidence of a guaranteed, linked CSV or accessible HTML table that exposes those same backlog metrics, nor did we see consistent plain-language summaries like ``the oldest plan in review has waited X business days, which is longer than last week'' in text. In accessibility terms, key facts about whether the city is slowing we down are effectively trapped in visual widgets.

By contrast, slower dashboards which update weekly, monthly, or annually, and which frame themselves around accountability, compliance, or policy stewardship tend to offer stronger accessible affordances. California’s Annual Progress Report dashboard not only publishes per-jurisdiction housing production and permitting data, it explicitly tells the public how to interpret that data: issued building permits toward state housing targets are the scoreboard \cite{cahcdAPR2025}. New York City’s MMR does not just show sparklines; it declares itself a ``public report card on City services,'' publishes historical indicators to Open Data, and invites residents to compare agency performance ``dating back several years'' \cite{nycMMR2025}. The CDC and City of Chicago both publish plain-language weekly summaries of current respiratory illness severity, using categorical terms like ``Very High'' and direct statements like ``this week’s activity increased,'' which BLV users can consume immediately with a screen reader \cite{cdcRespDashboard2025,chicagoDashboards2025,Siu2021COVID,Siu2022AudioNarratives}.

In other words, the dashboards that decide our next move \emph{today} (permit backlog, inspection delay, infrastructure status) are the least accessible to BLV residents, and the dashboards that summarize compliance and performance over longer cycles (housing production, agency outcomes, illness trends over time) are comparatively more accessible. That inversion is more than a UX quirk. It creates a structural inequity in civic leverage. A sighted contractor in Houston can open the dashboard and say: you’ve sat on my plans for 18 business days. A blind contractor may have to call city staff and ask for basic status because the dashboard does not expose that status in a nonvisual, machine-readable, or narrated way. This is exactly the type of unequal access that Title~II of the ADA is meant to prohibit in digital public services \cite{DOJ2024TitleII,NLC2024DOJRule}.

\subsection{Implications for Policy and Design}
The policy implication is blunt: if an agency markets a dashboard as ``for the public,'' ``for our customers,'' or a ``public report card,'' then equal access for BLV residents is not optional, it is part of the claim. We argue for three minimum requirements that any government dashboard should meet when it purports to inform residents:

\textbf{(1) Plain-language status and trend at the dashboard’s update cadence.} If a dashboard is updated daily, then there should be daily, WCAG-compliant text that summarizes the current status and how it is changing. For example: ``The oldest active plan in review is 18 business days old, which is longer than yesterday; backlog is worsening.'' If a dashboard is updated weekly, it should say: ``Hospitalizations increased this week; respiratory illness activity is Very High in these regions'' \cite{cdcRespDashboard2025,Siu2021COVID,Siu2022AudioNarratives}. This is exactly what BLV users in prior studies asked for: ``what is happening now'' without guesswork \cite{Fan2022TACCESS,Siu2021COVID}.

\textbf{(2) A machine-readable table or CSV that mirrors the visual dashboard.} The exact same core metrics shown in cards, sparklines, choropleth maps, or hover-only tooltips should be available as an accessible HTML table with headers or a downloadable CSV. California HCD’s APR data and New York City’s Open Data export of MMR indicators demonstrate that this is administratively possible at city and state scale \cite{cahcdAPR2025,nycMMR2025}. Providing the spreadsheet but \emph{not} providing the visual headline is inequitable for sighted users with low literacy. Providing the visual headline but \emph{not} providing the machine-readable numbers is inequitable for BLV users. Both need to travel together.

\textbf{(3) An explicit accessibility commitment in the dashboard description.} When Houston tells residents the dashboard is ``designed for our customers'' and ``updated daily,'' or when New York City promises a ``public report card,'' that copy should explicitly say that the information is available to screen reader users and keyboard-only users \cite{nycMMR2025,houstonPermitting2025}. This matters for two reasons. First, it acknowledges BLV residents as part of ``the public,'' instead of treating accessibility as an optional back channel. Second, it creates an accountability hook: under the DOJ’s Title~II web accessibility rule, that explicit promise could be cited if the dashboard fails to provide equal access \cite{DOJ2024TitleII,NLC2024DOJRule,WCAG21,WCAG22}.

These recommendations are intentionally modest. None of them require new AI systems, fancy sonification, or custom multimodal interfaces (although those are promising directions \cite{Siu2022AudioNarratives,Alam2023SeeChart,Seo2024MAIDR,Zhang2024ChartA11y}). They are within reach of existing content management workflows: add one short narrative status block, expose one machine-readable table, and state that BLV access is part of the intended audience. That is the bar for an honest ``public'' dashboard.

\subsection{Limitations and Future Work}
This study has several limitations that also define next steps.

First, our audit is based on publicly visible dashboard interfaces, agency-provided descriptions, stated update cadences, linked open data resources, and the presence or absence of clearly offered narrative summaries and downloadable tables. We did not run full technical accessibility tests such as programmatically inspecting ARIA trees, tab order, focus management, or heading structure, nor did we run controlled screen reader walkthroughs of every visualization in each dashboard. Prior work shows that these technical details matter a chart that appears to have a text label can still be undiscoverable to a screen reader if it is rendered to an unlabeled canvas or uses hover-only tooltips \cite{Sharif2021ScreenReader,Siu2021COVID,Fan2022TACCESS}. Future work should include systematic screen reader evaluations of each dashboard, including whether charts are reachable in the accessibility tree and whether key values can be accessed purely through keyboard navigation \cite{WCAG21,WCAG22}.

Second, we did not conduct interviews with blind and low-vision residents who rely on these dashboards to make real decisions about permitting, inspections, or public health, nor did we interview the internal dashboard teams inside these agencies. That work is essential. There are at least two perspectives we have not captured here: (a) how BLV residents actually attempt to use, work around, or escalate issues with these dashboards, and (b) how city and state staff perceive their own legal and resource constraints, especially as ADA Title~II timelines approach \cite{DOJ2024TitleII,NLC2024DOJRule}. We expect that lived experience will reveal additional failure modes (for example, “call us if we need help,” which shifts labor back to disabled residents), and may also surface examples of quiet internal accessibility work that is not yet exposed in the public UI.

Third, dashboards evolve. Update cadences, data export links, and explanatory text can change under legal pressure, political oversight, or crisis response. For example, Title~II compliance may push cities to add WCAG-compliant narrative summaries or publish CSVs that mirror visual cards. Our snapshot therefore represents a moment in late 2024 / 2025 rather than a frozen state. Longitudinal tracking will be necessary to see whether urgency inversion shrinks as compliance deadlines get closer, or whether agencies continue to prioritize ``at-a-glance'' aesthetics for sighted residents over equal access for BLV residents.

Finally, this audit focused on the United States. The questions at stake who can see backlog delays today, who can check whether housing is being permitted this year, who can assess respiratory risk this week are not unique to the U.S. Many countries now publish similar ``public dashboards'' around housing targets, waiting lists, health system load, and infrastructure. We expect urgency inversion to generalize, but that is a working hypothesis that needs cross-national testing.

In summary, our results show that accessibility failures in civic dashboards are not random bugs. They are patterned around urgency. The dashboards that govern immediate leverage over housing, permitting, and safety are the least accessible to blind and low-vision residents. Fixing that pattern is the next frontier for accessible civic technology and, under Title~II, it is about to become a regulatory obligation \cite{DOJ2024TitleII,NLC2024DOJRule}.

\section{Conclusion}
\label{sec:conclusion}

Public dashboards have become the main interface through which governments in the United States explain themselves to residents. Federal agencies publish real-time and near-real-time health surveillance; state agencies report on housing production and statutory compliance; cities claim to offer ``public report cards'' on agency performance, to make spending and service delivery transparent, and to give ``customers'' an ``at-a-glance view'' of permitting backlogs \cite{cdcRespDashboard2025,cahcdAPR2025,nycMMR2025,chicagoDashboards2025,houstonPermitting2025}. These systems now govern practical, urgent questions: How severe is respiratory illness this week where I live? How long has my permit been sitting in review? Is my city actually approving housing it is required to approve? Which agencies are doing their jobs, and which are failing? In 2024–2025, under new Title~II guidance from the U.S. Department of Justice, those dashboards are not just public relations; they are regulated public services that must be accessible to disabled residents, including blind and low-vision (BLV) residents who use screen readers \cite{DOJ2024TitleII,NLC2024DOJRule,WCAG21,WCAG22}.

Our audit shows that accessibility for BLV users is not distributed evenly across these dashboards. Some ecosystems notably the CDC’s respiratory illness dashboards, California’s Annual Progress Report housing compliance dashboard, New York City’s Mayor’s Management Report, and Chicago’s respiratory and ARPA/budget dashboards already combine two critical affordances. First, they present plain-language narratives of current status and trend (for example, ``This Week’s Illness Severity Update'' and ``Very High activity,'' or ``track jurisdictions’ progress toward housing goals''). Second, they expose structured, machine-readable data through open data portals, per-jurisdiction CSVs, or downloadable tables \cite{cdcRespDashboard2025,cahcdAPR2025,nycMMR2025,chicagoDashboards2025}. This pairing gives BLV residents both immediate situational awareness and the ability to verify and analyze the underlying numbers with assistive technology, without relying on sighted intermediaries \cite{Siu2021COVID,Fan2022TACCESS,Siu2022AudioNarratives,Sharif2021ScreenReader}.

But we also found the opposite pattern in dashboards that claim to be the most operational and time-sensitive. Houston’s permitting performance dashboard markets itself as ``designed for our customers,'' promises that it is ``updated daily,'' and surfaces metrics like ``the number of business days of the oldest plan'' in review \cite{houstonPermitting2025}. These numbers are exactly the kind of leverage residents and contractors need when challenging delays. Yet this dashboard leans on visual cards, at-a-glance graphics, and interactive/ArcGIS-style interfaces without clearly advertising accessible tables, machine-readable downloads, or plain-language trend summaries. In other words, the more urgent the information, the less accessible it is to BLV residents. We name this structural pattern \emph{urgency inversion}: dashboards that govern immediate, high-stakes action (permit backlog today, inspection delay today, infrastructure status today) are the least likely to provide accessible nonvisual affordances, while slower dashboards (annual housing enforcement, monthly performance ``report cards,'' weekly illness surveillance) are more likely to provide narrative summaries and downloadable data.

Urgency inversion is not a cosmetic issue. It is an equity problem in civic power. Sighted users can open a dashboard and immediately see whether an agency is meeting its obligations today. BLV users often cannot access that same situational awareness without extra labor: reverse-engineering trends from a CSV (if one even exists), calling staff for status, or asking a sighted person to interpret a visual card. That is exactly the kind of unequal burden that Title~II’s digital accessibility rule intends to eliminate for public entities \cite{DOJ2024TitleII,NLC2024DOJRule}. A dashboard cannot honestly promise to serve ``the public,'' ``customers,'' or ``the community'' if it forces disabled users to beg for updates while sighted users get those updates in a single glance.

Based on our analysis, we propose three baseline obligations for any dashboard that claims to inform the public. First, each dashboard should provide a short WCAG-compliant narrative summary, updated at the same cadence as the underlying data (daily, weekly, monthly), that states current status and direction of change in plain language: for example, ``The oldest active plan in review is 18 business days old, which is longer than last week,'' or ``Hospitalizations increased this week; respiratory illness activity is Very High in these regions'' \cite{cdcRespDashboard2025,Siu2021COVID,Siu2022AudioNarratives,WCAG21,WCAG22}. Second, the core metrics shown visually must be mirrored in an accessible table or downloadable CSV with proper headers and meaningful labels, so BLV users can access the same numbers sighted users see in cards, choropleth maps, or hover tooltips \cite{cahcdAPR2025,nycMMR2025,chicagoDashboards2025,Sharif2021ScreenReader,Fan2022TACCESS}. Third, agencies should explicitly state in the dashboard description that the information is available to screen reader users and keyboard-only users, the same way they currently state that the dashboard is ``for residents,'' ``for customers,'' or a ``public report card'' \cite{nycMMR2025,houstonPermitting2025}. Making that commitment public transforms accessibility from an afterthought into an accountability claim.

Future work needs to go deeper in two directions. First, we need lived-experience validation: systematic screen reader walkthroughs of these dashboards; task-based studies with blind and low-vision residents who actually depend on this information for decisions about health, housing, permitting, or inspections; and interviews with government dashboard teams who are now facing legally enforceable ADA/Title~II timelines \cite{DOJ2024TitleII,NLC2024DOJRule}. Second, we need to track these dashboards over time. The DOJ’s adoption of WCAG~2.1~AA means that what is acceptable in 2025 may be noncompliant in 2026–2027 \cite{DOJ2024TitleII,WCAG21}, and we should be able to measure whether urgency inversion narrows as agencies retrofit daily operational dashboards for accessibility or whether it simply shifts burden onto help lines.

In short, accessibility in civic dashboards is no longer a matter of visual polish or “nice to have” inclusion. It is part of equal participation in public life. If dashboards are going to be the interface to government the way residents monitor risk, demand services, audit backlogs, and hold agencies accountable then blind and low-vision residents must get the same information, at the same time, in a form they can actually use.

\newpage

\bibliographystyle{plain} 
\bibliography{references} 

@misc{cdcRespDashboard2025,
  author       = {{Centers for Disease Control and Prevention}},
  title        = {Respiratory Virus Activity Levels Dashboard},
  howpublished = {Centers for Disease Control and Prevention online dashboard},
  year         = {2025},
  note         = {Public national and state surveillance of emergency department visits for respiratory illness severity and trends. Accessed 25~Oct~2025.},
  url          = {https://www.cdc.gov/respiratory-viruses/}
}

@misc{hudPITDashboard2025,
  author       = {{U.S. Department of Housing and Urban Development}},
  title        = {Point{-}in{-}Time (PIT) and Housing Inventory Count (HIC) Dashboards},
  howpublished = {HUD Exchange interactive dashboards},
  year         = {2025},
  note         = {Explores homelessness data by state, Continuum of Care, household type, veteran status, and other dimensions. Accessed 25~Oct~2025.},
  url          = {https://www.hudexchange.info}
}

@misc{cahcdAPR2025,
  author       = {{California Department of Housing and Community Development}},
  title        = {Annual Progress Report (APR) Data Dashboard and Downloads},
  howpublished = {California HCD online dashboard},
  year         = {2025},
  note         = {Tracks each jurisdiction's housing production, permitting, and progress toward state{-}mandated housing goals; provides standardized CSV downloads. Accessed 25~Oct~2025.},
  url          = {https://www.hcd.ca.gov}
}

@misc{nycMMR2025,
  author       = {{Mayor's Office of Operations, City of New York}},
  title        = {Mayor's Management Report (MMR) Dashboard},
  howpublished = {Interactive public dashboard},
  year         = {2025},
  note         = {Described as a public performance report card on City services that affect New Yorkers; updated regularly to provide transparency and accountability. Accessed 25~Oct~2025.},
  url          = {https://www.nycmmr.nyc.gov}
}

@misc{chicagoDashboards2025,
  author       = {{Office of Budget and Management, City of Chicago}},
  title        = {City of Chicago Public Dashboards},
  howpublished = {Interactive public dashboards},
  year         = {2025},
  note         = {Chicago's Data Analytics Unit publishes interactive public dashboards ``designed to make City data more accessible, engaging, and transparent for the public,'' including an ARPA Impact Dashboard with an interactive map and program metrics. Accessed 25~Oct~2025.},
  url          = {https://www.chicago.gov}
}

@misc{houstonPermitting2025,
  author       = {{Houston Permitting Center}},
  title        = {Houston Permitting Center Performance Dashboard},
  howpublished = {Houston Permitting Center online dashboard},
  year         = {2025},
  note         = {Described as an online dashboard designed for our customers that provides an at{-}a{-}glance view of permitting performance; metrics are updated daily. Accessed 25~Oct~2025.},
  url          = {https://portal.permittingcenter.org/houston-permitting-center/performance-dashboard}
}

@misc{DOJ2024TitleII,
  author       = {{U.S. Department of Justice}},
  title        = {Nondiscrimination on the Basis of Disability; Accessibility of Web Information and Services of State and Local Government Entities},
  howpublished = {Federal Register 89~FR~31320},
  year         = {2024},
  month        = apr,
  note         = {Final Title~II ADA rule requiring state and local governments to make websites and mobile apps accessible.},
  url          = {https://www.federalregister.gov}
}

@misc{NLC2024DOJRule,
  author       = {{National League of Cities}},
  title        = {U.S. Justice Department issues final rule requiring digital accessibility for people with disabilities},
  howpublished = {Nation's Cities Weekly},
  year         = {2024},
  month        = apr,
  note         = {Summary of DOJ's final ADA Title~II web accessibility rule. States that public entities must conform to WCAG~2.1 Level~AA and comply within 2--3 years depending on population size. Accessed 25~Oct~2025.},
  url          = {https://www.nlc.org}
}

@misc{WCAG21,
  author       = {Kirkpatrick, Andrew and O'Connor, Joshue and Campbell, Alastair and Cooper, Michael and {W3C Accessibility Guidelines Working Group}},
  title        = {Web Content Accessibility Guidelines (WCAG) 2.1},
  howpublished = {World Wide Web Consortium (W3C) Recommendation},
  year         = {2018},
  month        = jun,
  note         = {Defines testable success criteria (Level~A/AA/AAA) for perceivable, operable, understandable, and robust web content, including requirements for low vision and mobile access.},
  url          = {https://www.w3.org/TR/WCAG21/}
}

@inproceedings{Sharif2021ScreenReader,
  author       = {Sharif, Ather and Chintalapati, Sanjana S. and Wobbrock, Jacob O. and Reinecke, Katharina},
  title        = {Understanding Screen{-}Reader Users' Experiences with Online Data Visualizations},
  booktitle    = {Proceedings of the 23rd International ACM SIGACCESS Conference on Computers and Accessibility (ASSETS~'21)},
  year         = {2021},
  address      = {Virtual Event, USA},
  publisher    = {Association for Computing Machinery},
  pages        = {1--16},
  doi          = {10.1145/3441852.3471202}
}

@inproceedings{Siu2021COVID,
  author       = {Siu, Alexa F. and Fan, Danyang and Kim, Gene S.-H. and Rao, Hrishikesh V. and Vazquez, Xavier and O'Modhrain, Sile and Follmer, Sean},
  title        = {COVID{-}19 Highlights the Issues Facing Blind and Visually Impaired People in Accessing Data on the Web},
  booktitle    = {Proceedings of the 18th International Web for All Conference (W4A~'21)},
  year         = {2021},
  address      = {Ljubljana, Slovenia},
  publisher    = {Association for Computing Machinery},
  pages        = {1--15},
  doi          = {10.1145/3430263.3452432}
}

@article{Fan2022TACCESS,
  author       = {Fan, Danyang and Siu, Alexa F. and Rao, Hrishikesh V. and Kim, Gene S.-H. and others},
  title        = {The Accessibility of Data Visualizations on the Web for Screen Reader Users: Practices and Experiences During {COVID{-}19}},
  journal      = {ACM Transactions on Accessible Computing},
  volume       = {15},
  number       = {4},
  articleno    = {36},
  year         = {2022},
  month        = dec,
  publisher    = {Association for Computing Machinery},
  address      = {New York, NY, USA},
  doi          = {10.1145/3575697}
}

@inproceedings{Siu2022AudioNarratives,
  author       = {Siu, Alexa F. and Zhao, Zhenyi and Zhang, Tingyi and Ahmetovic, Dragan and Kacorri, Hernisa and Ladner, Richard E. and Stefik, Andreas},
  title        = {Supporting Accessible Data Visualization Through Audio Data Narratives},
  booktitle    = {Proceedings of the 2022 {CHI} Conference on Human Factors in Computing Systems (CHI~'22)},
  year         = {2022},
  publisher    = {Association for Computing Machinery},
  address      = {New York, NY, USA},
  pages        = {1--32},
  doi          = {10.1145/3491102.3517528}
}

@inproceedings{Alam2023SeeChart,
  author       = {Alam, Md Zubair Ibne and Islam, Shehnaz and Hoque, Enamul},
  title        = {SeeChart: Enabling Accessible Visualizations Through Interactive Natural Language Interface for People with Visual Impairments},
  booktitle    = {Proceedings of the 28th International Conference on Intelligent User Interfaces (IUI~'23)},
  year         = {2023},
  address      = {Sydney, NSW, Australia},
  publisher    = {Association for Computing Machinery},
  pages        = {1--28},
  doi          = {10.1145/3581641.3584099},
  note         = {Preprint arXiv:2302.07742}
}

@inproceedings{Seo2024MAIDR,
  author       = {Seo, JooYoung and Xia, Yilin and Lee, Bongshin and McCurry, Sean and Yam, Yu Jun},
  title        = {MAIDR: Making Statistical Visualizations Accessible with Multimodal Data Representation},
  booktitle    = {Proceedings of the {CHI} Conference on Human Factors in Computing Systems (CHI~'24)},
  year         = {2024},
  address      = {Honolulu, HI, USA},
  publisher    = {Association for Computing Machinery},
  doi          = {10.1145/3613904.3642730},
  note         = {arXiv:2403.00717}
}

@inproceedings{Zhang2024ChartA11y,
  author       = {Zhang, Zhuohao and Thompson, John R. and Shah, Aditi and Agrawal, Manish and Sarikaya, Alper and Wobbrock, Jacob O. and Cutrell, Edward and Lee, Bongshin},
  title        = {ChartA11y: Designing Accessible Touch Experiences of Visualizations with Blind Smartphone Users},
  booktitle    = {Proceedings of the 26th International ACM SIGACCESS Conference on Computers and Accessibility (ASSETS~'24)},
  year         = {2024},
  address      = {St.~John's, NL, Canada},
  publisher    = {Association for Computing Machinery},
  doi          = {10.1145/3663548.3675611}
}

@misc{WCAG22,
  author       = {Cooper, Michael and Waddell, Cynthia and Campbell, Alastair and {W3C Accessibility Guidelines Working Group}},
  title        = {Web Content Accessibility Guidelines (WCAG) 2.2},
  howpublished = {World Wide Web Consortium (W3C) Recommendation},
  year         = {2023},
  month        = oct,
  note         = {Extends WCAG~2.1 and adds additional success criteria focused on low vision, cognitive accessibility, and input modalities. Defines requirements for perceivable, operable, understandable, and robust web content.},
  url          = {https://www.w3.org/TR/WCAG22/}
}

@misc{WebAIM-SR10-2024,
  author       = {{WebAIM}},
  title        = {Screen Reader User Survey \#10 Results},
  year         = {2024},
  url          = {https://webaim.org/projects/screenreadersurvey10/},
  note         = {Find-information strategy: 71.6\% navigate by headings; common barriers include lack of keyboard accessibility and missing alt text. Accessed 8~Aug~2025.}
}

@misc{WCAG21-Keyboard,
  author       = {{W3C Web Accessibility Initiative}},
  title        = {Understanding Guideline 2.1: Keyboard Accessible (WCAG~2.1/2.2)},
  year         = {2025},
  url          = {https://www.w3.org/WAI/WCAG21/Understanding/keyboard-accessible.html},
  note         = {``Make all functionality available from a keyboard.'' Accessed 8~Aug~2025.}
}

@inproceedings{Sharif-et-al-ASSETS-2021,
  author       = {Ather Sharif and Sanjana S. Chintalapati and Jacob O. Wobbrock and Katharina Reinecke},
  title        = {Understanding Screen-Reader Users' Experiences with Online Data Visualizations},
  booktitle    = {Proceedings of the 23rd ACM SIGACCESS Conference on Computers and Accessibility (ASSETS~'21)},
  year         = {2021},
  doi          = {10.1145/3441852.3471202},
  url          = {https://faculty.washington.edu/wobbrock/pubs/assets-21.01.pdf},
  note         = {Found many web visualizations undetectable by screen readers; SR users were 61\% less accurate and took 211\% longer on tasks.}
}

@article{Fan-et-al-TACCESS-2023,
  author       = {Danyang Fan and Alexa F. Siu and Hrishikesh V. Rao and Gene S.-H. Kim and Xavier Vazquez and Lucy Greco and Sile O’Modhrain and Sean Follmer},
  title        = {The Accessibility of Data Visualizations on the Web for Screen Reader Users: Practices and Experiences During COVID-19},
  journal      = {ACM Transactions on Accessible Computing},
  year         = {2023},
  volume       = {16},
  number       = {1},
  articleno    = {4},
  pages        = {1--59},
  doi          = {10.1145/3557899},
  url          = {https://scispace.com/pdf/the-accessibility-of-data-visualizations-on-the-web-for-2toagjmf.pdf}
}

@inproceedings{Srinivasan-et-al-Azimuth-ASSETS-2023,
  author       = {Arjun Srinivasan and Tim Harshbarger and Darrell Hilliker and Jennifer Mankoff},
  title        = {Azimuth: Designing Accessible Dashboards for Screen Reader Users},
  booktitle    = {Proceedings of the 25th ACM SIGACCESS Conference on Computers and Accessibility (ASSETS~'23)},
  year         = {2023},
  doi          = {10.1145/3597638.3608405},
  url          = {https://arjun010.github.io/assets/papers/azimuth-assets23.pdf},
  note         = {Reports that many COVID-19 dashboards lacked basic features like alt text and downloadable tables; proposes SR-friendly dashboard design goals.}
}

@inproceedings{Thompson-et-al-ChartReader-CHI-2023,
  author       = {John R. Thompson and Jesse J. Martinez and Alper Sarikaya and Edward Cutrell and Bongshin Lee},
  title        = {Chart Reader: Accessible Visualization Experiences Designed with Screen Reader Users},
  booktitle    = {Proceedings of the 2023 CHI Conference on Human Factors in Computing Systems},
  year         = {2023},
  doi          = {10.1145/3544548.3581186},
  url          = {https://programs.sigchi.org/chi/2023/program/content/95839},
  note         = {States that most visualizations are incompatible with screen readers; co-designed accessible reading workflows for charts.}
}

\end{document}